\documentclass[12pt,preprint]{aastex} %
\usepackage{graphics}

\begin{document}
 
\shorttitle{ Balmer Lines as Diagnostics
of Physical Conditions in AGN Broad Emission Line Regions}

\shortauthors{L. \v C. Popovi\'c} 
 
\title{Balmer Lines as Diagnostics
of Physical Conditions in AGN Broad Emission Line Regions
}

\author{L. \v C. Popovi\'c\altaffilmark{1,2,3}}
 
\altaffiltext{1}{Astronomical Observatory,  Volgina 7, 11160 Belgrade
74, Serbia}

\altaffiltext{2}{
Astrophysikalisches Institut Potsdam, An der Sternwarte 1, D-14482 
Potsdam, Germany }

\altaffiltext{3}{Isaac Newton Institute of Chile,
 Yugoslavia Branch}
 \email{lpopovic@aob.bg.ac.yu, lpopovic@aip.de}

\begin{abstract}
Using a well known method for laboratory plasma diagnostic,  the 
Boltzmann-plot, we discuss the physical properties
in Broad Line Region (BLR) of Active Galactic Nuclei 
(AGN).  We apply the  Boltzmann-plot 
method to Balmer lines  on a sample 
of 14 AGN,  
finding that  it may indicate the 
existence of "Case B" recombination or Partial Local Thermodynamical 
Equilibrium (PLTE).
For BLR of AGN, where PLTE exists, we estimated the 
electron temperature and density of BLR. The estimated electron 
temperature ($T\sim 13000 - 37000$ K) are in good agreement 
with previous 
estimates. The estimated electron densities depend on opacity of the 
emitting plasma in BLR. They are from $N_e\sim 10^9\rm cm^{-3}$ for 
optically thick to  $N_e\sim 10^{14}\rm cm^{-3}$ for optically thin 
emission plasma in BLR. The estimated electron temperature has 
been shown to be velocity dependent, and it decreases for 
higher velocities.
Although  the alternative explanation than PLTE indicated by 
Boltzmann-plot  may be considered (e.g. high intrinsic reddening), 
the method may give the quick estimate of physical 
conditions in BLR before to apply the sophisticated methods.
 \end{abstract}

%\keywords{galaxies:  Seyfert -- line: profiles -- spectroscopy} 

\section{Introduction}  

The 
emission line spectrum of Active Galactic Nuclei (AGN) is produced over a 
 wide range of distances from the  central continuum source, and 
under a wide 
 range of physical and kinematical conditions  (see Sulentic et al. 
2000, and 
references therein). 
 The line strengths, their widths and shapes are the powerful 
tools
for emitting gas diagnostics in  different parts of the emitting 
region of an AGN. The physics in the Broad Line Region (BLR) is 
more complicated 
than in the
Narrow Line Region (NLR).
The classical and  recent studies point toward photoionization as the main 
heating source for  the BLR emitting gas (see e.g. Kwan 
\& Krolik 1981,  Osterbrock 1989, Baldwin et al. 1995, Marziani et al. 1996, 
Baldwin et al. 1996, Ferland et al. 1998, Krolik 1999).
The photoionization, recombination and collisions can be considered 
as relevant 
processes in BLRs. At larger ionization parameters, recombination is 
more important, but at the higher temperatures the collisional excitation 
become also important as well as in the case of low ionization 
parameters. 
These two effects, as well as radiative-transfer effects in Balmer lines 
should be taken into account to explain the
 ratios of Hydrogen lines  (Osterbrock 1989, Krolik 1999). 

Different types of the physical conditions and processes can be 
assumed in order to use the emission lines for diagnostic of emission plasma 
(Osterbrock 1989, Griem 1997, Ferland et al. 1998). Although "in 
many 
aspects the BLRs are physically as  closely related to stellar atmospheres 
as traditional nebula" (Osterbrock  1989), the plasma in the  BLR 
probably  
does  not come close to being in complete Local Thermodynamical 
Equilibrium (LTE).
 However, there may still be the Partial Local Thermodynamical 
Equilibrium (PLTE) 
in the sense that  populations of sufficiently highly excited levels are 
related to the next  ion's ground state population by Saha-Boltzmann 
relations (van der Mullen  et al. 1994), or to the total
population in all fine-structure levels of the ground-state configuration
(see Griem 1968, 1997). The PLTE for different types of plasma: ionizing, 
recombining plasma and plasma in ionization balance were discussed in 
Fujimoto \& McWhirter (1990). They found that the populations of 
higher-lying levels are well described by the Saha-Boltzmann equation. On 
the other hand van der Mullen et al. (1994) confirmed that "if electrons 
realize an equilibrium between ionization and recombination in a 
two-temperature radiationless plasma, then the Sahal equation can be 
obtained by replacing the thermodynamic temperature by the electron one".  
Moreover, recently Popovi\'c et al. (2002) found  that the Balmer lines of 
NGC 3516 indicate that the Balmer emitting line  region  may be in the PLTE.

The aim of this paper is to test   the existence of PLTE,  and discuss the 
possibility of estimation of the relevant  physical processes and plasma 
parameters  in BLRs using the Boltzmann-plot of Balmer lines.

\section{Theoretical remarks}

As a simple case let us consider the optically thin plasma.
In the case of 
 plasma of the length $\ell$ along the line of sight, 
spectrally
integrated emission line intensity (I$_{lu}$) is given as (see e.g. Griem 
1997,
Konjevi\'c 1999)

$$I_{lu}={hc\over\lambda}g_{u}A_{ul}\int_0^\ell
N_udx\approx {hc\over\lambda}A_{ul}g_u\ell{N_0\over 
Z}\exp(-E_u/kT_e),\eqno(1)$$
where $\lambda$ is transition wavelength, $g_u$  statistical weight of 
the upper level, $A_{ul}$  transition probability, $Z$  the partition 
function, $N_0$   the total number density  of radiating species, 
$E_{u}$  the energy of the 
upper level, $T_e$  electron temperature and $h,\ c,\ k$ are the well 
known constants (Planck, speed of light and Boltzmann 
constant, respectively).

If the plasma is in PLTE, the population 
of the parent energy  states adhere to a Boltzmann distribution 
uniquely 
characterized by 
their excitation temperature ($T_e$ in Eq. (1)), and this temperature may 
be 
obtained from a Boltzmann-plot when the transitions within the same 
spectral series are considered

 $$\log(I_n)=\log{I_{ul}\cdot
\lambda\over{g_uA_{ul}}}=B-A{E_u},\eqno(2)$$
where $I_{lu}$ is relative intensity of transition from upper to lower 
level ($u\to l$), $B$ and $A$ 
are constants, where $A$ indicates temperature and we will 
call it the temperature parameter. 

If we can approximate the $\log(I_n)$ as a linear decreasing function 
of 
$E_u$ 
then: a) it indicates that PLTE may exist  at least to some extent  
in the BLR;
b) if  PLTE is present, the
population adhering to a Boltzmann distribution is uniquely characterized
by its excitation temperature. Then we can estimate the electron
temperature from Eq. (2), $T_e=1/(kA)$, where $k=8.6171\cdot 10^{-5}\rm 
eV/K$ is the Boltzmann constant;
c) if PLTE is present  we can roughly estimate the minimal electron 
density in BLR.
 Here, we should mention that "Case B" recombination 
of Balmer lines  can bring the $\log(I_n)$ {\it vs} $E_u$ 
as linear decreasing function (Osterbrock 1989). But, regarding the physical 
conditions 
(electron densities and temperatures) in BLRs in this case 
 the constant A is too small ($A<0.2$) and the Boltzmann-plot cannot 
be 
applied 
for diagnostics of electron temperature even if PLTE exists  (see discussion in Sec. 4). 
Moreover, in this case Boltzmann-plot method can be used as an 
indicator 
of "Case B" recombination  in BLRs of some AGN.

 Taking into account that the Balmer lines originate from the same series,
we can use the Boltzmann-plot relation for testing the existence of 
PLTE or 
"Case B" recombination. Also, here we should mention that we will assume 
that $\ell$, the length  of  the Balmer line formation region, is the same 
for all Balmer lines.

\section{Observations,  data reductions and measurements}

In order to test the existence of PLTE in BLR,
we use HST observations obtained with the Space Telescope Imaging
Spectrograph (STIS) and Faint Object Spectrograph (FOS),  covering the 
wavelength ranges  2900-5700 \AA\ 
and
6295-6867 \AA\  (rest wavelength). 
 From the
very large data base of AGN spectra at HST archive we selected the objects 
using following selection criteria:  
 a)  the observation covered the Balmer series line wavelength region;  b)  
the observations were performed on the same day; 
c)  all the lines from Balmer series can be recognized and all have 
relatively well defined shapes; 
d) we considered only low red-shifted objects. 
The list 
of 14 selected objects is presented in Table 1.

 The spectra were reduced by the HST
team.  We transformed the wavelength scale to zero red-shift  taking into
account the cosmological red-shift of the objects (V\'eron-Cetty \& 
V\'eron 
2000). After that we estimated and subtracted the continuum. Estimated 
error 
of  5\% - 10\%
due to subtraction of the continuum is included in the cumulative 
error (see Figs.1-3). The fluxes of the lines were measured by using the 
DIPSO software. The measured flux ratios of broad Balmer lines are present 
in Table 2.

\subsection{Satellite and narrow lines}

To perform a test we  subtracted the narrow and  satellite lines from 
Balmer lines. To estimate the contribution of these lines we used 
a multi-Gaussian analysis.
 We fit each line with a sum of Gaussian components using a
$\chi^2$ minimalization routine (Figs. 1-3).

To limit the number of free parameters in the fit of the H$\beta$ line 
we
have also set some {\it a priori} restrictions on the narrow components
and satellite lines (Popovi\'c et al.
2001,2002). In the first place, the three narrow Gaussians representing
the
two [OIII]$\lambda\lambda 4959,5007$ lines and the narrow H$\beta$
component are preconditioned to have the same red-shift and full widths
proportional to their wavelengths.  In the second place, we have linked
the
intensity ratio of  the two [OIII] lines according to  the atomic value
(line strength),
1:3.03.  Finally, we have included in the fit a shelf of  Fe II 
template (Korista 1992, Popovi\'c et al. 2001,2002). An 
example of fitting H$\beta$ line region of WPV 007 is present 
in Fig. 1, the dashed lines are subtracted from H$\beta$ 
profile.

In the case of H$\alpha$, we { assume} that the [NII]$\lambda\lambda 
6548,6583$ and the H$\alpha$
narrow components have the same red-shift and full widths proportional 
to
their wavelengths (dashed lines at the bottom of the Fig. 2). Taking 
into
account that the two [NII] lines belong to the transition between the
same multiplet we { assume} that their intensity ratio is 1:2.96 (see
e.g. Wiese et al. 1966).

In the case of an AGN with strong
narrow lines we estimated  the contribution of [OIII]4363 line in 
H$\gamma$ flux using the  ratio of  (I$_{4959}+$I$_{5007})/$I$_{4363}$ 
lines adopting that $T\sim 10000$ K (see Table 11.5 in Osterbrock 
1989) and low electron density characteristic for NLRs. Where it was 
possible, we also fit H$\gamma$ line assuming that [OIII]4363 line can 
be represented with one Gaussian having the same full width as [OIII] 
lines proportional to its wavelength  (see Fig. 3).
  The 
contribution of the [NeIII]3967 line 
in H$\varepsilon$ has been neglected. 

The error-bars, presented in Figs. 5, are estimated as cumulative
errors
due
to  continuum subtraction error ($\sim$10\%), and error of the line
flux measurements.  
One  should note that we approximatively subtracted narrow and 
satellite lines,  but we
estimated that error in this procedure was within the frame of errors
presented on the graphs.

\subsection{Reddening}

On the other hand, the reddening effect can influence  the Balmer lines 
ratio (see e.g. Crenshaw \& Kraemer 2001, Crenshaw et al. 2001, 2002, and 
references therein) and consequently on temperature parameter obtained by 
Boltzmann-plot. Here  the  Galactic reddening was taken into account 
using the data from NASA's Extragalacitc Database (NED). 
  In order to 
test the total (Galactic + intrinsic) reddening influence we have 
considered the case of Akn 564,
 where the reddening data are given by Crenshaw et al. (2002). We 
estimated that the reddening effect can contribute to the Boltzmann-plot 
slope 
around of 30\%-40\% (in the case of Akn 564 around of 35\%, 
$(kA)/(kA)_{redd.}=1.35$), but this effect cannot 
qualitatively 
disturb the straight line as a function in Boltzmann-plot. 
The  reddening effect will 
always cause that  temperatures measured by this method will give
smaller values if  this effect is not  taken into 
account.  
In the rest of our sample we  
did not consider the intrinsic reddening 
effect.

\subsection{Velocity measurements}

The data quality required for a comparison of temperature parameter 
$A$ as a 
function of velocity is quite high. The data quality are different for 
different spectra of AGN. Consequently, for velocity measurements we used 
only $H\alpha$ and $H\beta$ lines.
 To investigate the velocity obtained on Full 
Width 
at Half Maximum (FWHM) and Full Width at Zero Intensity (FWZI) as a 
function  of temperature for considered AGN, we first 
normalized the 
cleaned profiles of these lines to one and converted wavelength 
axis to the  $X=(\lambda-\lambda_0)/\lambda_0$ scale (see Fig. 4). 
In order to investigate the random velocity dependence which is 
probably 
related to physical conditions (in this case electron temperature) we 
measured FWHM and FWZI of an averaged line profile obtained from 
 $H\alpha$ and $H\beta$ profiles.  The 
differences between FWHM and 
FWZI of $H\alpha$ and $H\beta$ were relatively small, except in NGC 
1566.

\section{Results and discussion}

In Figs. 5-7 and in Table 2 we present our results. As one can see 
from 
Figs. 
5 and 6, the Boltzmann-plot of the Broad Balmer Lines indicates the 
existence 
of 
PLTE at least in some 
parts of BLR of significant fraction of the considered AGN (mainly 
BLRG).
If we take into account that intrinsic reddening can amplify the constant 
$A$  then we have  three cases: i) Boltzmann-plot cannot be applied  
 at all (e.g. NGC 1566), ii) Boltzmann-plot can be applied, but the 
constant 
is small $A<0.3$, then even if the PLTE exists, the ratio of Balmer 
line fluxes 
are nearly independent of temperature. This case can rather be 
treated 
as (or 
indicates) recombination. iii) Boltzmann-plot can be applied and
$A>0.3$. In 
this 
case we can consider that PLTE exists in the sense that the populations 
of 
the higher-lying levels ($n \ge 3$) are well described by the 
Sahal-Boltzmann equation, and excited temperature can be replaced by 
electron temperature.

 From the 14 selected AGN, we can say that in 9 AGN
Boltzmann-plot indicates the existence of PLTE  ($A>0.3$, $T<40000$ 
K) in BLR, 
while in the case of  4 of them   (see Table 2 and Fig. 6) the 
Boltzmann-plot 
indicates 
"Case B" recombination in BLR. In the remaining 1 AGN (NGC 1566) the 
Boltzmann-plot cannot be applied (see Fig. 5).
Using this method we can determine the electron temperature.
In a favorable situation in laboratory, with this method the electron 
temperature may be determined
to 2-3\%, otherwise normally an uncertainty of about 5-10\% must be
expected for laboratory plasma (Konjevi\'c 1999). In our case,
 the determination of the temperatures is very sensitive to the flux
measurements. As one can see from Fig. 5  uncertainties in the
measured flux are higher for weaker lines. Moreover, the intrinsic 
reddening effect 
can increase the temperature.
 Consequently,
the  uncertainties in the temperature determination in this case
are around 30\%. 
 Valid criteria for PLTE
and criteria for the application of this spectroscopic method are widely 
discussed by Fujimoto \& McWhirter (1990), Griem (1997) and Konjevi\'c 
(1999). The measured flux ratio of Balmer lines and measured flux of 
H$\beta$ line and temperature parameter $A$ are given in Table 2.
As one can see from Table 2 there is no strong dependence between 
$F_{H\alpha}/F_{H\beta}$ ratio and temperature parameter. On the 
other hand, one can 
see from Table 2 and Figs. 6ab the temperature difference between  radio
quiet 
(RQ) and radio loud (RL) 
objects. Although this follows the fact  that RL and RQ AGN
have different distributions
in Balmer line widths (e.g. Wills and Browne 1986),
temperature differences RL and RQ AGN might be a result from selection
biases  rather than a difference in the radio properties. From Table 2
and Fig 6ab, one can conclude that in considered
sample of RL AGN PLTE might exist in BLRs.
 Here we should mention that an alternative to the PLTE may be a 
very high 
intrinsic reddening 
in this type of AGN. In any case the electron 
temperatures  are in good agreement with some previous estimates 
(see, e.g. 
Osterbrock 1989, Sulentic et al. 2000). 

To estimate the electron density we will consider two cases: optically thin 
and optically thick plasma.

For spatially homogeneous and inhomogeneous plasma, as
well as in a steady state plasma (applicable to slowly time-varying
plasma) but {\it optically thin plasma}, the method may  be used to 
estimate the lower 
limit of electron density (Griem 1997) taking that  $N_e\sim 7\cdot 
10^{18}z^6n_c^{-17/2} \sqrt{kT_e\over{E_H}}\rm \ cm^{-3}$,
where $z$ is the charge "seen" by the optical electron ($z=1$ for neutral
emitters), $n_c$ is the principal quantum number of
the upper level,  and E$_H$ is ionization energy of Hydrogen.
Taking $n_c=3$ (for the 
H$\alpha$ emission in PLTE),
the low limit of the electron density is of
the order of $\sim10^{13}-10^{14}\rm cm^{-3}$ (see also discussion
and Fig. 7.4. in Griem 1997). 

On the other hand, we cannot consider pure optically thin plasma in 
BLRs, 
and one should consider the case of {\it optically thick plasma} in 
BLR. For 
optically thick plasma, the equation of radiation transfer should be 
solved simultaneously with the rate equations for the population densities. 
However, sometimes it is adequate to assume that a reduced 
transition probability for the spontaneous transition describes the effect 
of reabsorption of the line. Within this approximation, one can take that
in Eq. (1) effective transition probabilities 
decrease by a factor $N_{esc}$, that is the mean number of scattering 
before escape which depends of the optical depth (see Eq. 12.6  in 
Osterbrock 
1989). Assuming that for all lines from Balmer spectral series $N_{esc}$ 
is similar, we still 
can use the Eq. (2), but in this case  
the value of critical $N_e$ should be reduced (Osterbrock 1989) 
$N_e'=N_e/N_{esc}$. Considering the critical electron density one may 
take 
the values for the H$\alpha$ optical depth  ($\tau (H\alpha)\sim 64 - 
4.5\cdot 10^4$) given in Osterbrock (1989). For these optical depths 
we obtain the 
values of electron density within the range from $N_e\sim
(10^9-10^{12})\rm 
cm^{-3}$. It is in agreement with the previous estimates 
(Osterbrock 
1989, Krolik 1999).

 The electron density estimated for optically thin plasma,  $\sim
10^{14}\ \rm cm^{-3}$,  is significantly higher than in some previous
estimates for the BLR ($10^{9}\ \rm cm^{-3}$ see, e.g. Osterbrock 1989, 
Sulentic et al. 2000). However, in some previous works a high electron 
density has been
also suggested (van Groningen 1987,  Sivron \& Tsuruta 1993, Brotherton
et al. 1994). But, for optical thick plasma the estimated electron 
densities are in agreement with conventionally accepted for BLRs.

On the other hand, it has been shown that the Balmer line ratios are
velocity 
dependent in AGN (Stirpe 1990,1991) and this is probably related to 
both a 
range of physical conditions (electron temperature and density) and  to the 
radiative transfer 
effects. Although, we have relatively small number of observed AGN for
 serious statistical analysis, we analyzed the temperature parameter 
as a 
function of random velocities at FWHM and FWZI obtained from an 
averaged 
profile of H$\alpha$ and H$\beta$ lines (Fig. 6).
As one can see from Fig. 6 the temperature parameter A tends to 
increase
with velocities, especially in the case of velocities 
measured at FWZI. In this case the function A {\it vs} FWZI has 
a linear trend. Also,
 The difference in A {\it 
vs} FWHM (Fig. 6a) and A {\it vs} FWZI  (Fig. 6b)  indicates that BLR 
is complex and that 
physical conditions of regions which contribute to the line core and 
line wings are different.
For the AGN where $A>0.3$ and where PLTE may be present, we plot the 
estimated electron density as a function of velocities measured at 
FWHM (Fig. 7a) and FWZI (Fig. 7b). As one can see from Figs. 7ab, the 
electron temperature decrease with velocities. The solid lines in Fig. 
7ab present the best fit with a linear function $T_e=c-d\cdot V$, 
where V 
is velocity measured at FWHM and FWZI. For constants $c$ and $d$ we 
obtained: $c=(44.0\pm 4)\cdot 10^3$ and $d=1.14\pm0.25$ for 
velocities measured at FWZI; and $c=(37.0\pm 3)\cdot 10^3$ and 
$d=1.87\pm0.50$ for velocities measured at  FWHM.

\section{Conclusion}
  
 Using the fact that the Balmer lines belong to the
 same spectral series, we apply the  Boltzmann-plot method to test 
the
presence 
of PLTE in BLR and discuss the relevant physical processes  in a sample of 
14 AGN. 
From our test we can conclude:

1)  From the 14 selected AGN, we found that in 9 AGN
Boltzmann-plot indicates the existence of PLTE in BLR,
while in the case of 4  of them   the Boltzmann-plot indicates
"Case B" recombination in BLR. In remaining 1 AGN the Boltzmann-plot 
cannot
be applied.

2) The estimated BLR electron temperatures  using Boltzmann-plot  
where PLTE exists are in a range (1.3 - 3.7)$\cdot 10^{4}$ K (within 
30\% accuracy). They are in a good agreement with the previous 
estimations.

The electron densities in BLR have been considered for optically thin and 
optically thick plasma and we found that

i) For optically thin plasma, the electron density in the case of 
PLTE, at least in some parts of BLR, should be higher than conventionally 
accepted for BLR.

ii) For optically thick plasma, the electron density in the case of 
PLTE is in 
agreement with the conventionally accepted for BLR.

On the other hand, the electron temperatures estimated by using 
Boltzmann-plot tend to be velocity dependent as a linear decreasing 
function of random velocities measured at FWHM as well as at FWZI.

Although,  an alternative of PLTE in some AGN may be very high 
intrinsic reddening effect, the Boltzmann-plot method may be used 
for fast insight into  physical processes in BLR of an AGN prior 
applying more sophisticated 
physical models. 

For future investigation it is needed to do the decomposition of complex 
Balmer lines in order to find  the part where this method can be fully 
applied and investigate the velocity dependence of Balmer line ratios on 
a large sample of AGN.
  
\section{Acknowledgments}
This work was supported by the Ministry of
Science,
Technologies and Development of Serbia through the project 
``Astrophysical
Spectroscopy of
Extragalacitc Objects''. Also, the work was supported by Alexander von 
Humboldt Foundation through the program for foreign scholars. I 
thank Prof. S. Djeni\v ze, Prof. M.S. 
Dimitrijevi\'c and Prof. N. Konjevi\'c for useful discussion and 
suggestions. I would like to thank to the anonymous referee for the 
very 
 useful comments.

\clearpage

 \begin{table*}
\begin{center}
      \caption[]{The list of the selected  AGN.}
\begin{tabular}{|c|c|c|c|c|}
\hline
Name &Z&  Class& Obs. Date& Ins./grat. \\
\hline
AKN 564& 0.025 & Sy 1.8& May 24, 1996 &FOS/G400,G570\\
MARK 493 & 0.031&  Sy 1& Sep 4, 1996&FOS/G400,G570 \\
NGC 1566 &0.004&Sy 1&Feb 8, 1991&FOS/G400,G570\\
NGC 4151 &0.003&Sy 1.5&Feb 10,1998&FOS/G400,G570\\
PG 1116+215 &0.177&Sy 1&May 26, 1996&FOS/G570,G780\\
PG 1402+261 &0.164&Sy 1&Aug 25, 1996&FOS/G570,G780\\
PG 1626+554 &0.132&Sy 1&Nov 19,1996&FOS/G570,G780\\
WPV 007 &0.029&Sy 1&Jul 30, 1996&FOS/G400,G570\\
NGC 3227 &0.004&Sy 1.5&Feb 8, 2000&STIS/G430L,G750L\\
3C 120 & 0.033&BLRG&Sep 9, 2000&STIS/G430L,G750L\\
3C 390.3 & 0.056&BLRG&Sep 11, 2000&STIS/G430L,G750L\\
3C 445  &0.056&BLRG&Sep 27, 2000&STIS/G430L,G750L\\
3C 273 &0.158&BLRG&Jan 31, 1999&STIS/G430L,G750L\\
3C 382 &0.059& BLRG&Sep 12, 2000&STIS/G430L,G750L\\
\hline
\end{tabular}
\end{center}
\end{table*}

 \begin{table*}
\begin{center}
      \caption[]{The measured flux ratio of Balmer lines of 
considered   AGN and temperature parameter  A. The flux ratio of Akn 564 
Balmer lines is taken from Crenshaw et al. (2002).
}
\begin{tabular}{|c|c|c|c|c|c|c|}
\hline
Name &$F_{H\alpha}/F_{H\beta}$&$F_{H\gamma}/F_{H\beta}$& $F_{H\delta}/F_{H\beta}$& 
$F_{H\varepsilon}/F_{H\beta}$&$F_{H\beta}\rm\ (Ergs\ cm^{-2}s^{-1})$& 
A  
\\
\hline
AKN 564& 3.78$\pm$0.53 & 0.35$\pm$0.06& 0.20$\pm$0.03 
&0.13$\pm$0.02&(3.13$\pm$0.34)E-13&0.354\\
MARK 493 &2.45$\pm$0.16& 
0.40$\pm$0.04&0.22$\pm$0.04&0.14$\pm$0.01&(9.63$\pm$0.34)E-14&0.169\\
NGC 1566 &3.17$\pm$0.48&0.53$\pm$0.09&0.36$\pm$0.06&0.39$\pm$0.15& 
(7.29$\pm$0.97)E-14&--\\
NGC 4151 &3.40$\pm$0.32&0.41$\pm$0.06 
&0.19$\pm$0.01&0.11$\pm$0.02&(5.57$\pm$0.24)E-12&0.357\\
PG 1116+215 &2.94$\pm$0.74 
&0.48$\pm$0.12&0.22$\pm$0.08&0.087$\pm$0.015&(3.43$\pm$0.51)E-13&0.316\\
PG 1402+261 &  2.69$\pm$0.33&0.51$\pm$0.11&0.26$\pm$0.04 
&0.11$\pm$0.02&(1.61$\pm$0.14)E-13&0.201\\
PG 1626+554 &2.94$\pm$0.42&0.48$\pm$0.09& 
0.28$\pm$0.06&0.19$\pm$0.02&(1.77$\pm$0.13)E-13&0.122\\
WPV 007 &3.09$\pm$0.42& 
0.53$\pm$0.08&0.21$\pm$0.05&0.12$\pm$0.03& 
(9.99$\pm$0.63)E-14&0.263\\
NGC 3227 &4.39$\pm$0.31&0.41$\pm$ 
0.03&0.24$\pm$0.01&0.15$\pm$0.02&(4.71$\pm$0.13)E-13&0.338\\
3C 120 &4.47$\pm$0.42& 0.66$\pm$0.06&0.26$\pm$0.04 
&0.13$\pm$0.02&(7.46$\pm$0.31)E-13&0.335\\
3C 390.3 
&5.39$\pm$0.33&0.36$\pm$0.05&0.11$\pm$0.01&0.08$\pm$0.01 
&(2.30$\pm$0.11)E-13&0.671\\ 
3C 445  &6.91$\pm$0.55 
&0.42$\pm$0.05&0.24$\pm$0.04&0.15$\pm$0.02&(1.84$\pm$0.10)E-13&0.499\\
3C 273 &3.19$\pm$0.32&0.41$\pm$0.03 
&0.12$\pm$0.01&0.06$\pm$0.01&(1.54$\pm$0.10)E-12&0.522\\
3C 382 
&4.43$\pm$0.31&0.35$\pm$0.02&0.043$\pm$0.003 
&0.033$\pm$0.004&(5.23$\pm$0.25)E-13&0.903\\ \hline
\end{tabular}
\end{center}
\end{table*}
\clearpage

\begin{figure}
\resizebox{8.2cm}{!}{\includegraphics{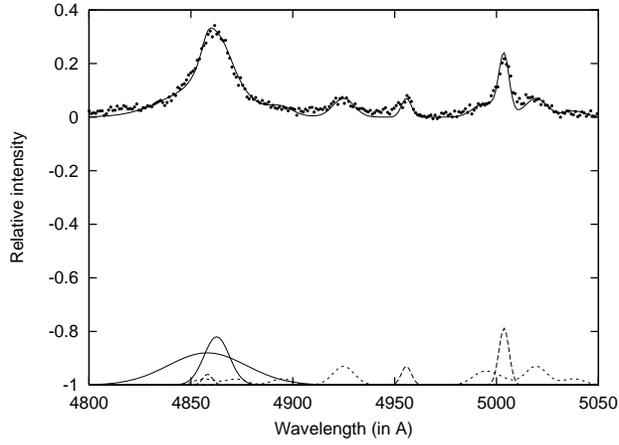}}
\caption{
Decomposition of H$\beta$ line of WPV007. The { dots represent} the
observation and solid line is the best fit. The Gaussian components are
shown at the bottom. The dashed lines at  bottom
represent the Fe II template, [OIII] and H$\beta$ narrow lines.}
\end{figure}

\begin{figure}
\resizebox{8.2cm}{!}{\includegraphics{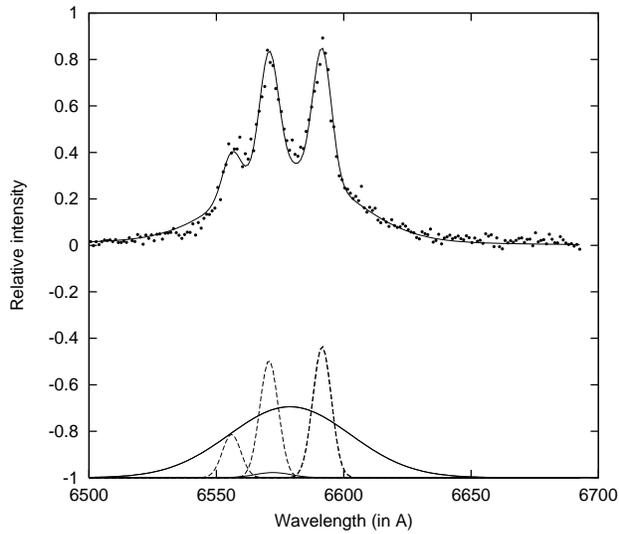}}
\caption{The same as in Fig 1, but for H$_\alpha$ of NGC 1566. The Gaussian 
components are
shown at the bottom. The dashed line at the bottom represent satellite N
II lines
and a narrow H$\alpha$ component.}
\end{figure}

\begin{figure}
\resizebox{8.2cm}{!}{\includegraphics{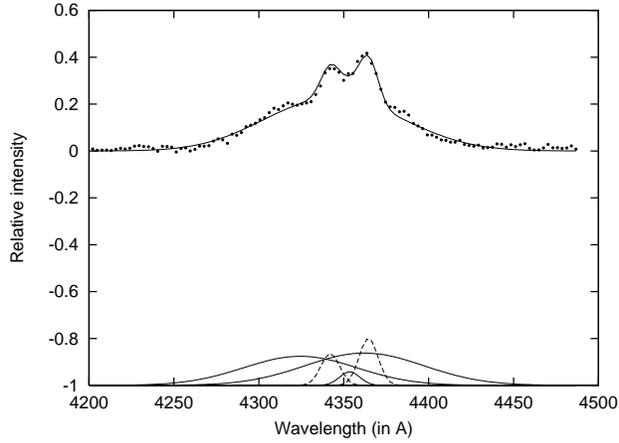}}
\caption{The same as in Fig 1, but for H$_\gamma$ of NGC 4151. The dashed 
line at the 
bottom represent satellite [OIII]4363 line
and a narrow H$\gamma$ component.}
\end{figure}

\begin{figure}
\resizebox{8.2cm}{!}{\includegraphics{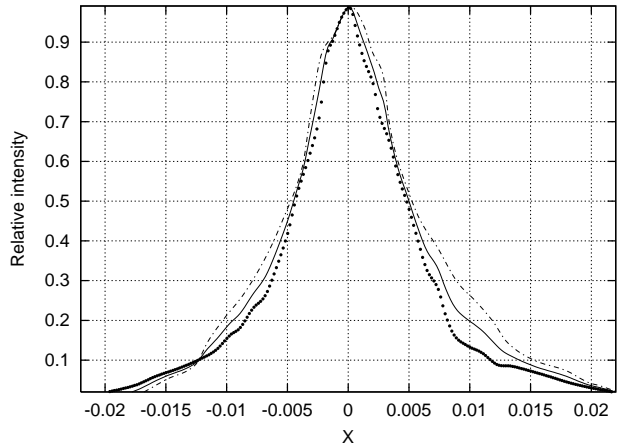}}
\caption{Comparison of the normalized profiles of PG 1116+215 H$\beta$ 
(dashed line) and 
H$\alpha$ lines (dots) with an averaged profile (solid line) as 
function of $X=(\lambda-\lambda_0)/\lambda_0$.}
\end{figure}

\begin{figure}
\resizebox{8.2cm}{!}{\includegraphics{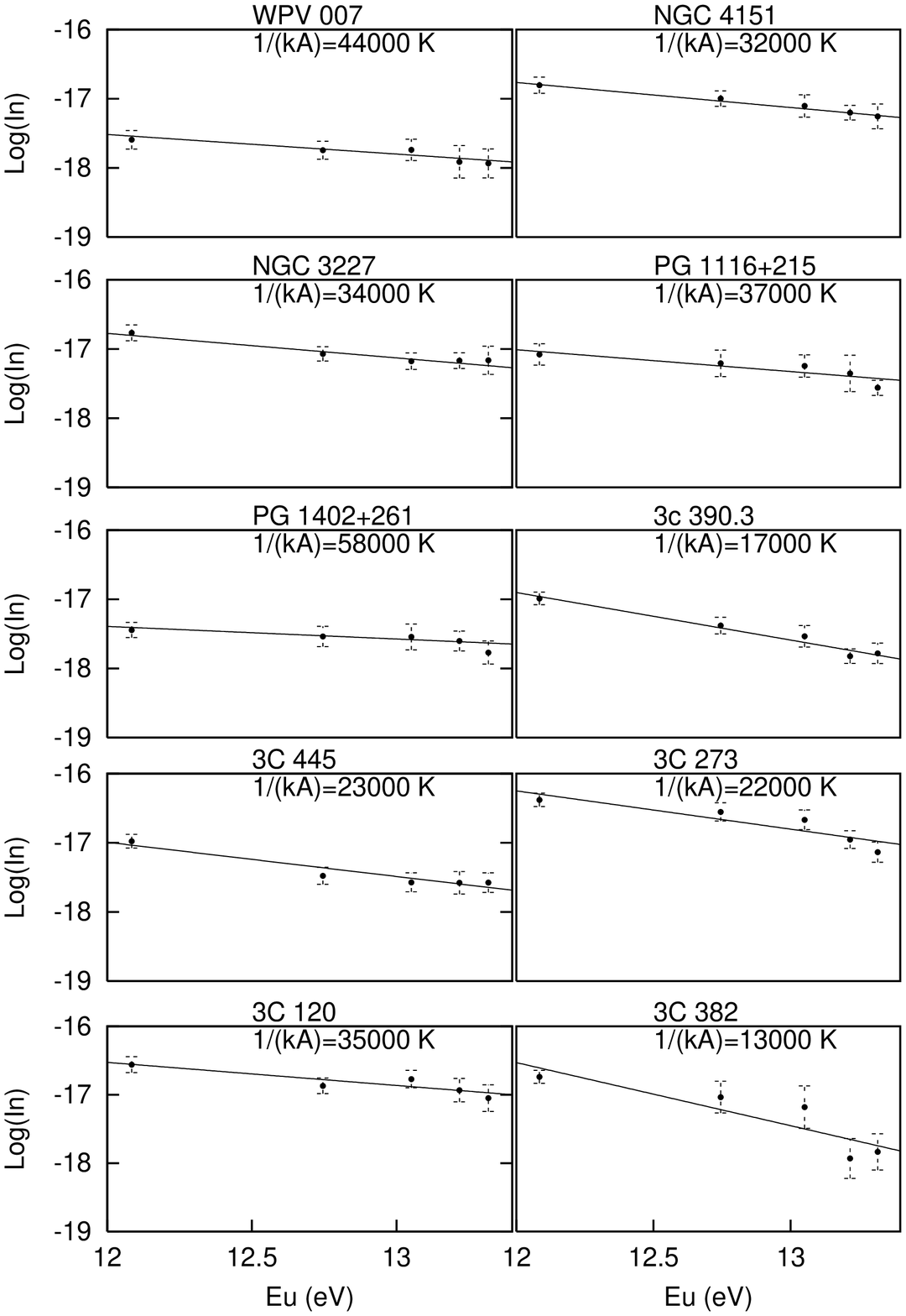}}
\resizebox{8.2cm}{!}{\includegraphics{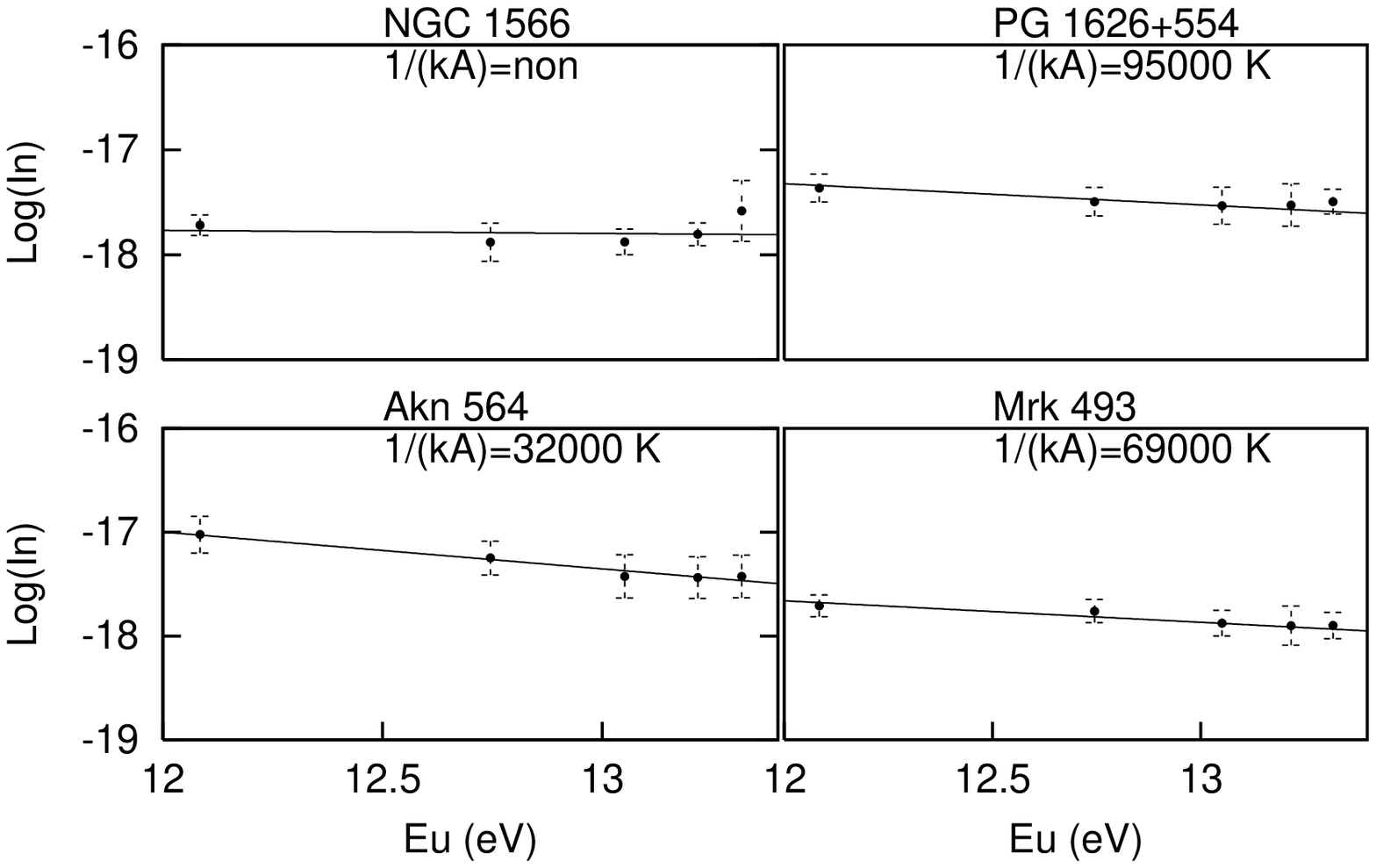}}
\caption{
The Boltzmann-plot for the  Balmer lines.
The corresponding temperature,
or $1/(kA)$, is given at the top of each panel. The $\log{I_n}$ of 
NGC 4151 is reduced on $\log{I_n}-1$}
\end{figure}

\begin{figure}
\resizebox{8.2cm}{!}{\includegraphics{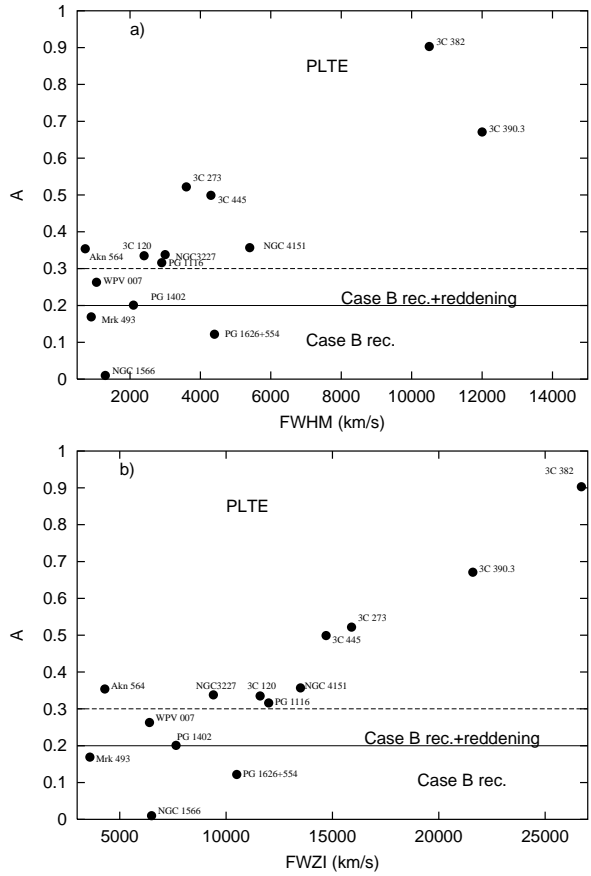}}
\caption{
The parameter $A$ obtained by Boltzmann-plot for all considered AGN 
as a 
function of random velocities measured at: a) Full Width 
at Half Maximum; b) Full Width at Zero Intensity. 
The values $A=0.2$ 
and $A=0.3$ are shown as a full and a dashed line, respectively.}
\end{figure}

\begin{figure}
\resizebox{8.2cm}{!}{\includegraphics{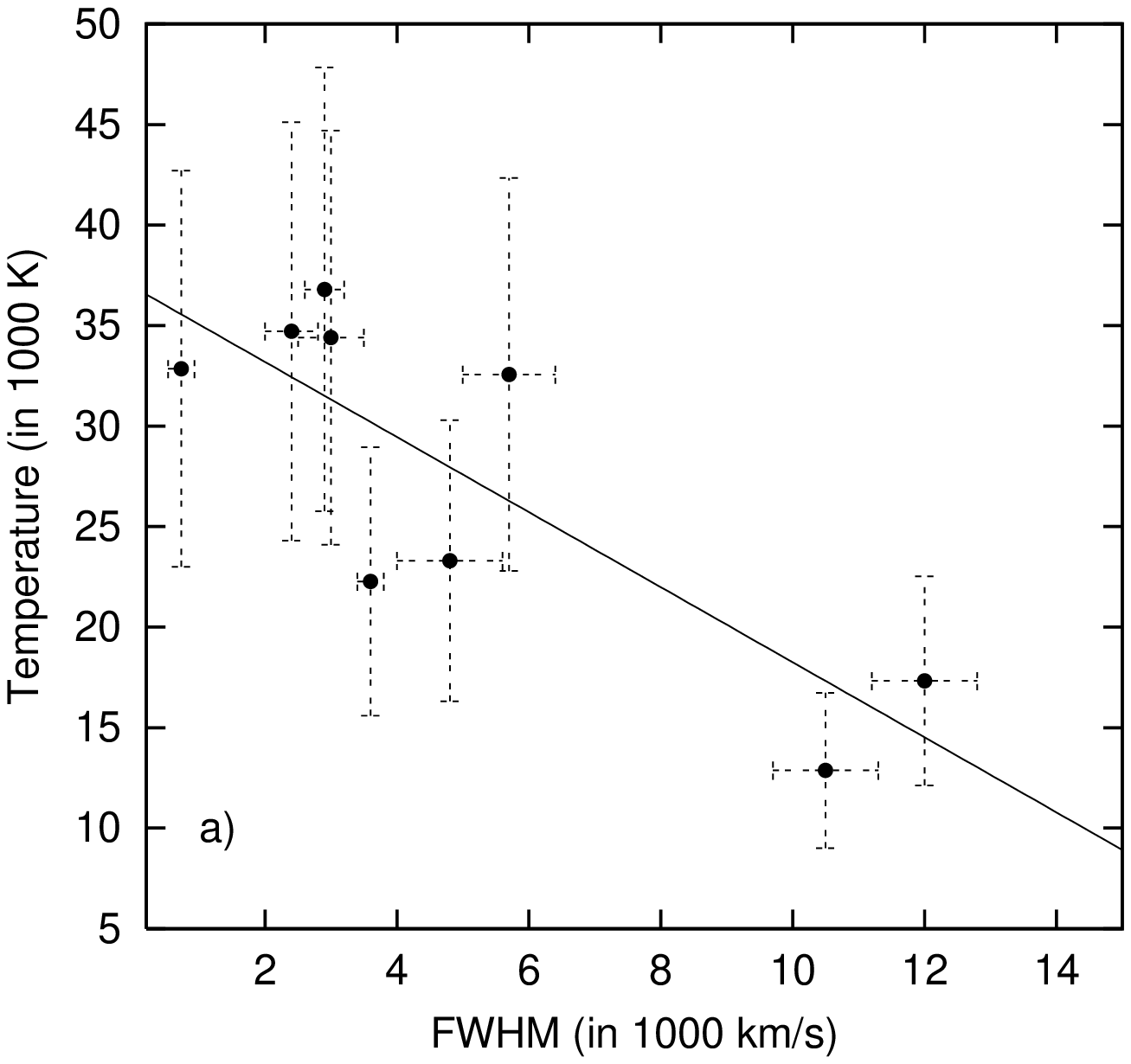}}
\resizebox{8.2cm}{!}{\includegraphics{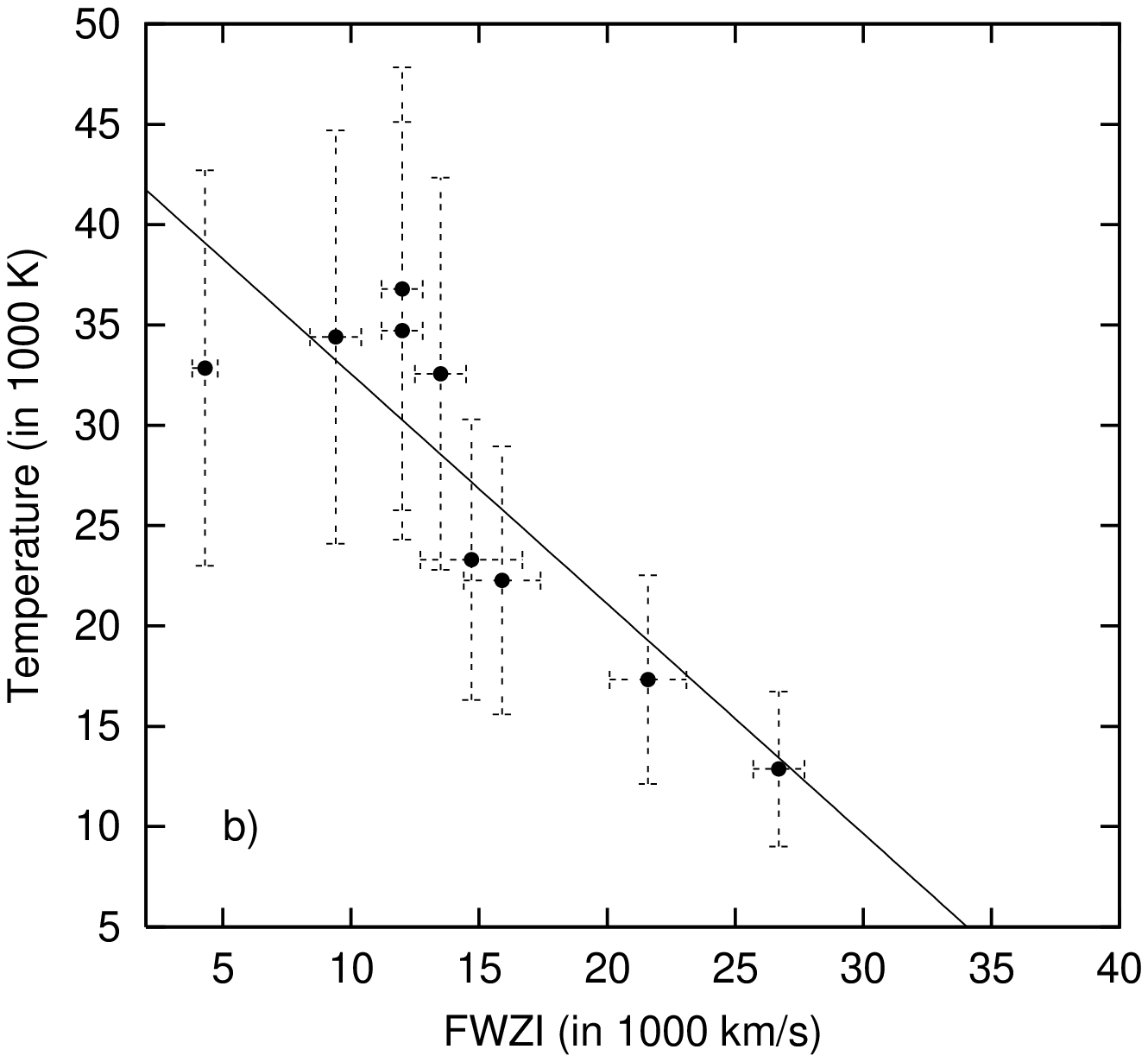}}
\caption{
The measured electron temperature using by Boltzmann-plot ($A>0.3$)  
as a function of 
velocities measured at: 
 a) Full Width
at Half Maximum; b) Full Width at Zero Intensity.}
\end{figure}

\end{document}